\documentclass[final,5p,times,twocolumn]{elsarticle}
\usepackage{epsfig}
\usepackage{ulem} 
\usepackage[usenames]{color}
\usepackage{graphicx}
\usepackage{amssymb}

\newcommand{\be}{\begin{eqnarray}}
\newcommand{\ee}{\end{eqnarray}}

\begin{document}

\begin{frontmatter}

\author[juli,uga]{M.~D\"oring}
\ead{m.doering@fz-juelich.de}
\author[uga,juli]{K.~Nakayama}\ead{nakayama@uga.edu}
\address[juli]{Institut f\"ur Kernphysik and J\"ulich Center for Hadron Physics, 
Forschungszentrum J\"ulich, D-52425 J\"ulich, Germany}
\address[uga]{Department of Physics and Astronomy, University of Georgia, Athens, GA 30602, USA}

\title{\hspace{14cm}{\tiny FZJ-IKP-TH-2009-29}
\\
On the cross section ratio $\sigma_n/\sigma_p$ in $\eta$ photoproduction}

\begin{abstract}
The recently discovered enhancement of $\eta$ photoproduction on the quasi-free neutron at energies around 
$\sqrt{s}\sim 1.67$ GeV is addressed within a SU(3) coupled channel model. The quasi-free cross sections on proton and neutron, $\sigma_n$ and $\sigma_p$, can be quantitatively explained. In this study, the main source for the peak in $\sigma_n/\sigma_p$ is a coupled channel effect in $S$ wave 
that explains the dip-bump structure in $\gamma n\to\eta n$. In particular, the photon coupling to the intermediate meson-baryon states is important. The stability of the result is extensively tested and consistency with several pion- and photon-induced reactions is ensured.
\end{abstract}

\begin{keyword}
Photoproduction of $\eta$ mesons \sep
multichannel scattering
\PACS 25.20.Lj\sep 
      13.60.Le\sep 
      14.40.Aq\sep 
      24.10.Eq\sep 
      14.20.Gk     
\end{keyword}

\end{frontmatter}


\section{Introduction}
The photo- and electroproduction of $\eta$ meson on the free proton has attracted extensive experimental effort in the last years~\cite{Krusche:1995nv,Dugger:2002ft,Ahrens:2003bp,Crede:2003ax,Nakabayashi:2006ut,Merkel:2007ig,Elsner:2007hm,Williams:2009yj,Crede:2009zz}. 
The prominent feature in these processes is the dominance of the $S$ wave contribution from threshold to energies beyond the $N^*(1650)$ region as revealed by various analyses~\cite{Tiator:1999gr,Chiang:2001as,Chiang:2002vq,Fix:2007st,Nakayama:2008tg,Anisovich:2008wd}. 

Recently, the reaction $\gamma n\to \eta n$ has become accessible in photoproduction experiments on the deuteron or nuclei~\cite{Kuznetsov:2007gr,Miyahara:2007zz,Mertens:2008np,Jaegle:2008ux,Anisovich:2008wd}. These measurements have been complemented by experimental studies of the beam asymmetry~\cite{Kuznetsov:2008hj,Kuznetsov:2008ii}. 
At energies around $\sqrt{s}\sim 1.67$ GeV, an excess of $\eta$ production on the neutron compared to the proton case has been reported~\cite{Kuznetsov:2007gr}; the result has been confirmed by other experiments~\cite{Miyahara:2007zz,Jaegle:2008ux}. This could be interpreted as a narrow nuclear resonance, but the interpretation is not unique~\cite{Anisovich:2008wd} and it is currently under an intense debate. Narrow nuclear resonances may also be accommodated in elastic $\pi N$ scattering at $1.68$ and $1.73$ GeV~\cite{Arndt:2003ga,Azimov:2003bb}. 

On the theoretical side, the structure observed in the quasi-free $\gamma n \to \eta n$ reaction has been interpreted as a potential signal for a non-strange member of an anti-decuplet of pentaquarks~\cite{Diakonov:1997mm,Polyakov:2003dx,Fix:2007st} (see also Refs.~\cite{Oh:2004,Diakonov:2004}, where a narrow baryon resonance has been suggested near 1.68 GeV). 

Within the framework of the Giessen model~\cite{Penner:2002md,Feuster:1998cj}, in Ref.~\cite{Shklyar:2006xw} the structure has been interpreted as an interference effect from the $S_{11}(1650)$ and $P_{11}(1710)$. In Ref.~\cite{Shyam:2008fr}, a subtle interference from various partial waves is made responsible for the observed structure for the $\eta n$ final state. In Ref.~\cite{Choi:2005ki}, an effective Lagrangian approach including an explicit narrow state was employed to describe the data of Ref.~\cite{Kuznetsov:2007gr}.
In the $\eta$-MAID~\cite{Chiang:2001as} analysis, the peak in $\sigma_n/\sigma_p$ is assigned to the $D_{15}(1675)$. This leads, however, to problems with too large an $\eta N$ decay width, and in Ref.~\cite{Fix:2007st} an additional narrow $P_{11}(1670)$ is considered.
In the analysis of Ref.~\cite{Anisovich:2008wd}, an interference within the $S_{11}$ partial wave alone has been found to give the most natural explanation. Also in Ref.~\cite{Miyahara:2007zz}, the $S_{11}$ assignment gives a much better fit to the data than that of $P_{11}$.

The findings described above have motivated us to study the $\eta$ photoproduction on proton and neutron within the $S$ wave model of Ref.~\cite{Doring:2009uc}.
This model, developed for the simultaneous description of $\gamma N\to\pi N$ and $\pi N\to\pi N$, can be easily extended to study $\eta N$, $K\Lambda$ and $K\Sigma$ final states which are included as coupled channels in the formalism. For details of the model we refer to Ref.~\cite{Doring:2009uc}. There, the formulation of the model is kept general enough to accommodate the different final states included here. 
The hadronic interaction is mediated by the Weinberg-Tomozawa interaction in the lowest order chiral Lagrangian. The attraction in the $S_{11}$ partial wave leads, through the unitarization of the on-shell factorized potential in a Bethe-Salpeter equation, to the formation of a dynamically generated pole that can be identified with the $N^*(1535)$. This picture of the $N^*(1535)$ \cite{Kaiser:1995cy,Kaiser:1996js,CaroRamon:1999jf,Inoue:2001ip,Nieves:2001wt,Kolomeitsev:2003kt} is quite different from the quark model picture~\cite{Zhao:2000iz,Zhong:2007fx}. The model also contains explicit resonance states which account for the $N^*(1650)$ and a phenomenological almost energy independent background.

The hadronic part of the present model~\cite{Doring:2009uc} has been developed following the lines of Ref. \cite{Inoue:2001ip}. 
For the electromagnetic interaction, the photon couples to the meson and baryon components in the model. These meson and baryon pole terms (cf. Fig.~\ref{fig:diagrams}) are fixed from the transversality of these amplitudes (see Eqs.~(10-19) in Ref.~\cite{Doring:2009uc}). The implementation of the photon interaction follows Refs.~\cite{Haberzettl:1997jg,Haberzettl:2006bn,Nakayama:2008tg}. See also Refs. \cite{Borasoy:2005zg,Borasoy:2007ku} for a formulation where most of the approximations made in \cite{Doring:2009uc} are avoided. Bare photon couplings to the genuine states are also included in the present model. 

For the results of Sec. \ref{sec:results} on the quasi-free $p$ and $n$ in the deuteron, we use the impulse approximation, i.e. higher order effects such as (hadronic) double scattering~\cite{Doring:2004kt} are neglected. In order to account for the Fermi motion of the nucleon inside the deuteron, we follow the prescription of Ref.~\cite{Anisovich:2008wd} and
 fold the cross section for the free nucleon case with the momentum distribution of the nucleon inside the deuteron, where the participating nucleon is set on-the-mass-shell once the energy conservation allows the reaction to take place~\cite{jaegle_private}. The deuteron wave function is generated based on the Bonn potential~\cite{Machleidt:2000ge}.


\section{Results}
\label{sec:results}
The model of Ref. \cite{Doring:2009uc} has been applied to the reactions $\gamma N\to\pi N$ and $\pi N\to\pi N$. In this study, we include the corresponding $E_{0+}$ multipoles and $S$ wave amplitudes in the fit, but additionally take into account the reactions $\gamma p\to\eta p$, $\gamma n\to\eta n$, $\pi N\to\eta N$, $\gamma N\to KY$, and $\pi N\to KY$ where $Y=\Lambda,\,\Sigma$. The free parameters of the model of Ref. \cite{Doring:2009uc} have been refitted using the additional data. The resulting parameters are quite similar to those 
of Ref.~\cite{Doring:2009uc}. In this work, we focus on the issue of the structure observed in the quasi-free $\gamma n \to \eta n$ reaction as discussed in the Introduction. In Sec. \ref{sec:discu1} we comment on the results for other reactions relevant to the present discussion.

 \begin{figure}
\begin{center}
\includegraphics[width=0.46\textwidth]{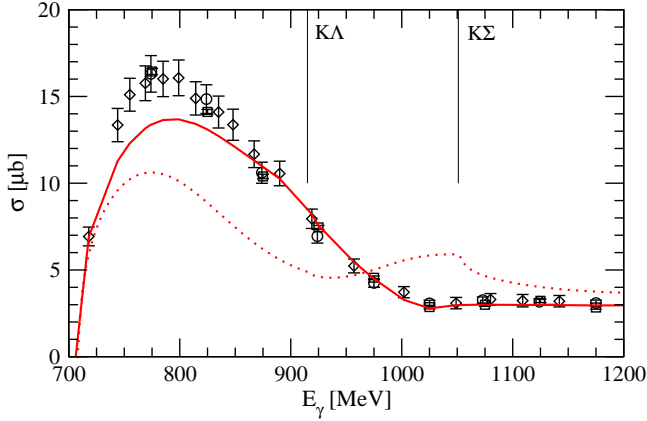}\\
\end{center}
\caption{The reaction $\gamma N\to\eta N$ on the free nucleon. Solid (dotted) line: present result for $\gamma p\to\eta p$ ($\gamma n\to\eta n$). The data are from JLab~\cite{Dugger:2002ft} (squares), Bonn~\cite{Crede:2003ax} (circles), and LNS~\cite{Nakabayashi:2006ut} (diamonds). Also, the $K\Lambda$ and $K\Sigma$ thresholds are indicated by the vertical lines.}
\label{fig:vaceta}
\end{figure}
In Fig. \ref{fig:vaceta} the result for the $\gamma p\to\eta p$ and $\gamma n\to\eta n$ cross sections on free nucleons is shown. The data for the proton case are well reproduced except for a slight under-prediction around the $N^*(1535)$ position. For the present study, a good data description above $E_\gamma=900$ MeV is essential, and this is indeed achieved. 
For the production on the free neutron, the cross section exhibits a minimum around $E_\gamma=930$ MeV, which is close to the $K\Lambda$ threshold, and a maximum at the $K\Sigma$ threshold. This dip-bump structure is absent for the proton case.

\begin{figure}
\begin{center}
\includegraphics[width=0.476\textwidth]{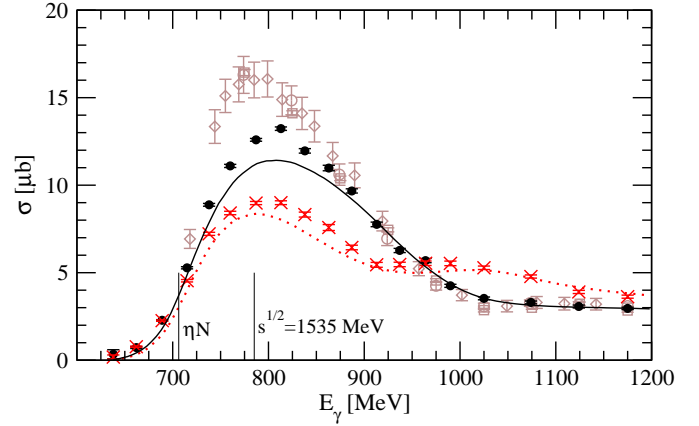}\\
\end{center}
\caption{Present result (Fermi folded) for the photoproduction on the quasi-free proton (solid line) and neutron (dotted line). The data are from Ref. \cite{Jaegle:2008ux} for the photoproduction on the quasi-free proton (solid circles) and neutron (crosses). The data for the free proton are also shown (open symbols, same as in Fig. \ref{fig:vaceta}). The vertical lines indicate the threshold energy of $\sqrt{s}=m_\eta+M_N$ for the free process and the nominal position of the $N^*(1535)$.}
\label{fig:deueta}
\end{figure}
In Fig. \ref{fig:deueta}, the present model is compared to the recent cross section data on the quasi-free neutron and proton in the deuteron from Ref. \cite{Jaegle:2008ux}.  The data are well reproduced.

\begin{figure}
\begin{center}
\includegraphics[width=0.46\textwidth]{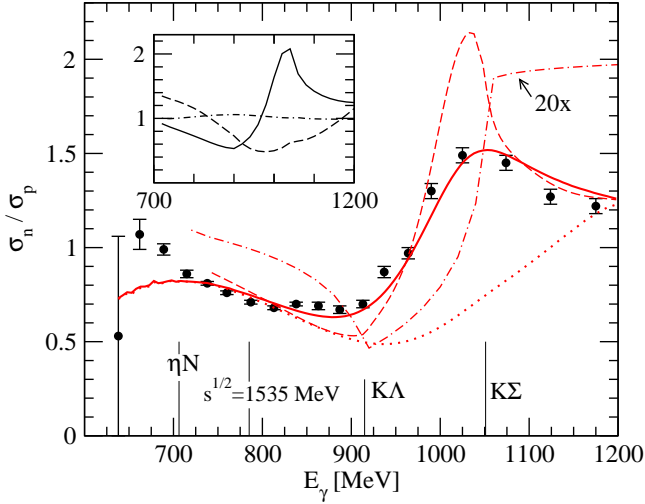}\\
\end{center}
\caption{The cross section ratio $\sigma(\gamma n\to\eta n)/\sigma(\gamma p\to\eta p)$ on the quasi-free nucleons in the deuteron. The data are from Ref. \cite{Jaegle:2008ux}. The full result is shown as the solid line; it includes the Fermi motion in the deuteron. The corresponding ratio in the case of free nucleons is shown as the dashed line. The dotted line shows the result after removing the $\gamma$ coupling to the $K^+\Lambda$ loop. Dash-dotted line: same as the dashed line but replacing the full final state interaction with the Weinberg-Tomozawa term. This curve is multiplied by an arbitrary factor of 20 to show the energy dependence of the ratio. Inset (free nucleon case): full result (solid line), w/o $\gamma$ coupling to $K\Lambda, \,K\Sigma$ (dashed line), only coupling to $\pi N$ intermediate state (dash-dotted line).}
\label{fig:ratio}
\end{figure}

In Fig. \ref{fig:ratio}, the ratio of cross sections of photoproduction on the quasi-free neutron over that on the quasi-free proton is shown (solid line). The data are from Ref. \cite{Jaegle:2008ux}. Earlier measurements~\cite{Weiss:2002tn,Hejny:1999iw,Krusche:2003ik} cover only the lower energy region but are in agreement with the new data of Ref. \cite{Jaegle:2008ux}. The dashed line in Fig. \ref{fig:ratio} indicates the ratio of cross sections on free nucleons, i.e., those shown in Fig. \ref{fig:vaceta}. The sharp structure in $\sigma_n/\sigma_p$ becomes Fermi smeared and the result for the quasi-free case (solid line) shows a broader peak in good agreement with the data.

\begin{figure}
\begin{center}
\includegraphics[width=0.43\textwidth]{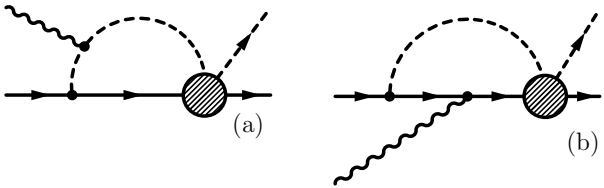}
\end{center}
\caption{Meson pole term (a) [$t$ channel] and (subleading) baryon pole term (b) [$u$ channel] in photoproduction. The hatched circles represent the unitary $MB\to\eta N$ amplitude. The contribution from the Kroll-Ruderman $\gamma N\to MB$ term arises automatically~\cite{Doring:2009uc}.}
\label{fig:diagrams}
\end{figure}

The appearance of the sharp peak in $\sigma_n/\sigma_p$ is obviously due to the
intermediate strangeness states in the model as indicated in Fig. \ref{fig:vaceta}. The difference in the cross sections on $p$ and $n$ arises from the isospin breaking of the photon couplings in the final state interaction loop contributions as illustrated schematically in Fig. \ref{fig:diagrams}.
For the meson pole loop (a), intermediate $\pi N$ and $K\Sigma$ states are possible in $\gamma n\to\eta n$, while in $\gamma p\to\eta p$, in addition, the $K^+\Lambda$ state is possible\footnote{The subleading contribution from the baryon pole term~\cite{Doring:2009uc} (cf. Fig.~\ref{fig:diagrams}(b)) is not included in the results, but discussed in Sec. \ref{sec:discu1}.}. 
Thus, there is a cancellation between the contribution from the intermediate $K^+\Lambda$ photon loop and from the other contributions ($\pi^+n$+$K^+\Sigma^0$ photon loops and terms with bare $\gamma NN^*$ couplings) in the $\gamma p \to\eta p$ reaction around $E_\gamma\sim 1.05$ GeV,  while this cancellation is absent in the $\gamma n \to\eta n$ reaction.

For the ratio $\sigma_n/\sigma_p$, the above discussed effect manifests itself in the observed peak structure in Fig.~\ref{fig:ratio}.
Indeed, removing the photon coupling to the $K^+\Lambda$ state in the $\gamma p \to\eta p$ reaction, one obtains the ratio given by the dotted line in Fig. \ref{fig:ratio}; the peak has disappeared.

In the following we discuss some further details of the underlying dynamics as well as the model dependence of the present results.
The inset in Fig. \ref{fig:ratio} shows again the free nucleon case (solid line). If, apart from the photon coupling to $K^+\Lambda$, we also remove the couplings to $K^+\Sigma^-$ (neutron case) and $K^+\Sigma^0$ (proton case), the dashed curve is obtained. If we remove additionally the bare photon couplings to the genuine states ($\gamma N\to N^*$), the only remaining photon coupling is to the $\pi^+ n$ (proton case) and $\pi^- p$ (neutron case) intermediate state. This is shown as the dash-dotted line in the inset. The ratio in this case is 1 over the entire energy range, up to tiny isospin breaking effects from the use of physical masses.

To check for the model dependence of the present results, we have replaced the hadronic final state interaction (FSI) with the Weinberg-Tomozawa (WT) term. For both reactions $\gamma p \to \eta p$ and $\gamma n \to \eta n$, the mechanism is then given by the loop graph (a) of Fig. \ref{fig:diagrams} with the unitary $MB\to \eta N$ transition (hatched circle) replaced by the WT term (This loop is sometimes referred to as ``triangle diagram'' in the literature). This parameter-free triangle graph is at order $1/f_\pi^3$ in the coupling and contributes at next-to-leading order (NLO) in the chiral expansion of the amplitude, as discussed in Sec. 3.2 of Ref.~\cite{Doring:2009uc} (cf. also Ref. \cite{Bernard:1994gm}). 
The resulting ratio $\sigma_n/\sigma_p$, shown as the dash-dotted line in Fig. \ref{fig:ratio}, is, of course, very different in magnitude from the full result (dashed line) --- replacing the strong, non-perturbative FSI by the tree-level WT term is certainly an oversimplification. Note in particular that, since the WT term does not provide direct $\pi N\to\eta N$ transitions, the otherwise large contribution from the $\pi N$ photon loop is absent in this case.
However, apart from the overall magnitude of the $\sigma_n/\sigma_p$ ratio, its energy dependence shows the same feature as the full result shown in Fig. \ref{fig:ratio}: in particular the slow fall-off around $E_\gamma\sim 800$ MeV is present, followed by the very steep rise when approaching the $K\Sigma$ threshold and the weak slope above the $K\Sigma$ threshold. 

Thus, on one hand a strong hadronic FSI interaction is needed to quantitatively explain the $\sigma_n$ and $\sigma_p$ cross sections in this non-perturbative energy region. On the other hand, the pronounced resonance-like enhancement of $\sigma_n$ compared to $\sigma_p$, at $E_\gamma\sim 1.05$ GeV, is already present when considering the triangle diagram at NLO in the chiral expansion. Note that this very triangle diagram (with $\pi^+n$ intermediate state) is also quantitatively responsible for the pronounced energy dependence of the cusp structure in near-threshold $\pi^0 p$ photoproduction~\cite{Doring:2009uc}~\footnote{There are, however, higher order terms not considered that induce a small, energy independent discrepancy for Re $E_{0+}(\pi^0p)$~\cite{Doring:2009uc}.}.


\section{Tests of $\eta$ photoproduction}
\label{sec:discu1}
\begin{figure}
\begin{center}
\includegraphics[width=0.46\textwidth]{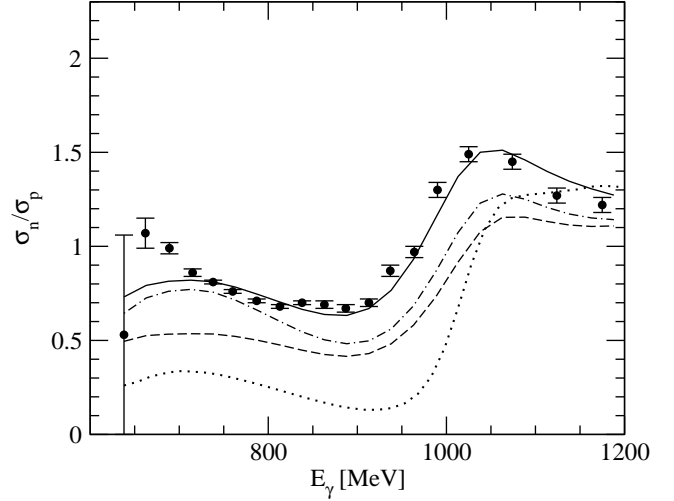}
\end{center}
\caption{Cross section ratio $\sigma(\gamma n\to\eta n)/\sigma(\gamma p\to\eta p)$ for the quasi-free processes. Solid line: full result, same as in Fig.~\ref{fig:ratio}.  Dotted line: Removing all contributions from genuine $N^*$ resonances.
Dash-dotted line: including the $\pi\pi N$ channel in the result. Dashed line: including the $\pi\pi N$ channel plus the baryon pole term (b) from Fig. \ref{fig:diagrams}.}
\label{fig:discussion}
\end{figure}
	
As discussed in the previous section, the final state interaction (FSI) from the unitarized $MB\to\eta N$ transition is needed for a quantitative description of the results. This FSI is strong and non-linear and thus it is difficult to fully disentangle the individual contributions. In particular, thresholds are present not only in the photon loops but also in the $MB\to\eta N$ amplitude, and for specific reaction channels, those thresholds may or may not appear pronounced. Still, the sensitivity of the results to changes of the model can be tested, which is done in the following. 

To start with, apart from the photon couplings to intermediate states discussed in the previous section, the current model~\cite{Doring:2009uc} contains explicit isospin breaking from the different bare couplings $\gamma p\to N^{*(+)}$ and $\gamma n\to N^{*(0)}$. To make sure these free parameters do not mock up the different cross sections in $\eta$ photoproduction on $p$ and $n$, we have removed all contributions from the two genuine resonances, i.e. the bare photon and strong couplings $\gamma N\to N^*$ and $MB\to N^*$ have been set to zero. 
The result is shown as the dotted line in Fig. \ref{fig:discussion}. The energy dependence of $\sigma_n/\sigma_p$
is barely changed but its magnitude is shifted downwards.
This is not unexpected, because one of the genuine resonances has its pole far in the complex plane and provides an almost energy independent background~\cite{Doring:2009uc}. Thus, removing this contribution results in the observed, almost energy independent shift of $\sigma_n/\sigma_p$ downwards. Second, this exercise tells us that the peak in Fig. \ref{fig:ratio} does not primarily come from the $N^*(1650)$ contribution.

We emphasize that this does not mean the $N^*(1650)$ plays no role in the current solution; a closer inspection of the $\sigma_n$ cross section shows a ``rounded cusp''~\cite{Baru:2004xg} at the $K\Sigma$ threshold, which can be a signal for a bound state with respect to the $K\Sigma$ channel. Here, this would be the $N^*(1650)$ resonance. Indeed, one needs this resonance for a quantitative description of the data shown in Figs.~\ref{fig:deueta} and \ref{fig:ratio} (and for the other reactions discussed below).

One can extract the resonance contributions to the amplitude in various ways. As argued in Ref. \cite{Doring:2009bi}, the most reliable and model independent procedure is the determination of pole positions and residues $a_{-1}$. We have performed this analysis for $\eta$ photoproduction on $p$ and $n$ using the current result. While the parameter set of the current solution is quite similar to that of Ref.~\cite{Doring:2009uc}, the pole position of the $N^*(1535)$ has changed from $1608-175\,i$ MeV to $1537-90\,i$ MeV. For the analysis of the photoproduction amplitude, also the photon loop from Fig. \ref{fig:diagrams}(a) has been analytically continued to the complex plane. However, it turns out that at the energies of interest of $\sqrt{s}\sim 1.67$ GeV, higher order terms in the Laurent expansion are important and the contributions from the resonance residues cannot saturate the amplitudes on $p$ and $n$ (although the energy dependence is matched quite well). This is a sign that resonances alone cannot explain the observed cross sections. This is in agreement with the finding mentioned above, where the genuine resonances have been removed and the peak position and size of $\sigma_n/\sigma_p$ have persisted.

Also, even if one evaluates $\sigma_n/\sigma_p$ using the original model from Ref.~\cite{Inoue:2001ip}, one still obtains a qualitatively similar result to that of the present study, although $\sigma_n$ and $\sigma_p$ are not well described individually at $E_\gamma\sim 1$ GeV.
The peak in $\sigma_n/\sigma_p$ is shifted, however, to somewhat lower energies of around $E_\gamma=900$ MeV. 
Note also that in Ref. \cite{Kaiser:1996js}, $\sigma_n/\sigma_p$ rises as a function of $E_\gamma$ as can be seen in Fig. 13 of Ref. \cite{Jido:2007sm}. 
Although the position of the sharp rise in $\sigma_n/\sigma_p$ can vary in different models, it seems to be a rather stable feature of various calculations within the chiral unitary framework. 

We have assumed that the total cross sections on $p$ and $n$ are dominated by the $S$-wave, the only partial wave included in this study. For the reaction $\gamma n\to\eta n$, the different partial wave analyses from Ref. \cite{Anisovich:2008wd} indicate that about 80 \% of the cross section arises from the $S$ wave in the energy region of $\sqrt{s}\sim 1.6$ to $1.7$ GeV. For the cross section on the proton, the situation is similar~\cite{Chiang:2001as,Nakayama:2008tg}. Thus, there is some but moderate change from higher partial waves, but assuming their energy dependence is smooth, those changes could be compensated by a refit of the current model, once these higher partial waves are taken into account.

Another question concerns the role of additional channels not included in the present model. We can estimate the influence of the $\pi\pi N$ channel by including it as in Refs.~\cite{Inoue:2001ip,Doring:2005bx}. It should be stressed, however, that the $\pi\pi N$ channel introduced in Ref.~\cite{Inoue:2001ip} was fitted to $\pi\pi N$ data only up to the $N^*(1535)$ resonance energy. Still, it gives a rough idea on how much this additional channel might alter the various pion- and photon-induced reactions described by the present model. The cross section in $\gamma p\to\eta p$ increases at the position of the $N^*(1535)$, once the $\pi\pi N$ channel is included. In fact, it then matches the data at the $N^*(1535)$ position, both for free and quasi-free protons. 
The ratio $\sigma_n/\sigma_p$ including the $\pi\pi N$ channel is shown as the dash-dotted line in Fig.~\ref{fig:discussion}; the influence of the $\pi\pi N$ channel is small for the ratio. 
So far, we have not considered the $\gamma$ coupling to the baryon component in the loop contribution as shown in Fig. \ref{fig:diagrams}(b). Although this process contributes little because it is subleading in the $1/M$ expansion~\cite{Jido:2007sm,Doring:2009uc}, it could significantly alter the ratio $\sigma_n/\sigma_p$, because the photon can couple to additional intermediate states like $\eta p$. However, the result for $\sigma_p$ and $\sigma_n$ does not change much with these higher order effects. The resulting ratio $\sigma_n/\sigma_p$ is shown as the dashed line in Fig. \ref{fig:discussion}.

There are other higher order effects like magnetic couplings or the $\Lambda\Sigma^0$ transition magnetic moment, all of them discussed in Ref. \cite{Jido:2007sm}. We expect moderate modifications of the results from these sources, but no qualitative changes, because they are small and partly cancel each other~\cite{Jido:2007sm}.


The consistency of the present model with the two independent multipoles $_nE_{0+}$ and $_pE_{0+}$ in $\pi$ photoproduction should be ensured in analogy with $\eta$ photoproduction on $n$ and $p$. The result of the present study is very similar to that shown in Ref.~\cite{Doring:2009uc} and the multipoles \cite{Drechsel:2007if,Arndt:2008zz} are well reproduced in the considered energy range of $E_\gamma\sim 700$ to $1200$ MeV. 
In particular, the phase of the $N^*(1535)$ is consistent with the analyses of Refs. \cite{Drechsel:2007if,Arndt:2008zz}. Note that the $\eta N$ threshold appears quite different in $_nE_{0+}$ and $_pE_{0+}$ (cf. Fig. 13 of Ref.~\cite{Doring:2009uc}). Just like in the present study of the $\eta n$ and $\eta p$ final states, it is the distinct photo-excitation of intermediate states that renders these two pion-multipoles so different.

As for the other reaction channels evaluated within the present model, we mention that a pronounced $K\Sigma$ threshold effect is observed in the $K\Lambda$ photoproduction on the proton. This is in contrast to the case of $\eta$ photoproduction, where this effect appears pronounced on  neutron but not on proton (cf. Fig. \ref{fig:vaceta}). Whether or not a threshold will appear pronounced depends on the specific reaction channel, due to the non-linearity of the amplitude.

We have also evaluated the other $S$-wave cross sections for the family of reactions $\pi N, \,\gamma N\to K Y,\,\eta N$ and compared to the corresponding data. While there is a qualitative overall agreement for most of these reactions, for some reactions the results are sensitive to the $\pi\pi N$ channel, which is not fully consistently included in the present model. This is beyond the scope of the present study because an inclusion of such channels as done e.g. in Refs. \cite{Sato:1996gk,Krehl:1999km,Penner:2002ma,Gasparyan:2003fp,Durand:2008es,Doring2} is difficult and requires the simultaneous study of all partial waves, beyond the $S$ wave as considered in this study. 


\section{Conclusions}
The excess of $\eta$ photoproduction on the neutron at photon energies around $E_\gamma\sim 1.02$ GeV ($\sqrt{s}\sim 1.67$ GeV), compared to the proton case, has been studied within a unitary coupled channel model that includes the SU(3) allowed meson-baryon channels $\pi N$, $\eta N$, $K\Lambda$, and $K\Sigma$. The photon couples to this hadronic system respecting gauge invariance as dictated by the Ward-Takahashi identity and unitarity. 

The experimentally determined $\eta$ photoproduction cross sections on quasi-free neutron and proton can be explained quantitatively within the present model which accounts for the $S$ wave contribution only. The intermediate $K\Sigma$ and $K\Lambda$ loop contributions to the $\gamma n \to n \eta$ and $\gamma p \to p \eta$ processes have been identified as the primary source of the difference in the observed energy dependence in these two reactions which, in turn, leads to the occurrence of a relatively sharp spike-like structure in the corresponding cross section ratio $\sigma_n/\sigma_p$. We emphasize that this is a direct consequence of the underlying hadron dynamics which, in the present model, is driven by the Weinberg-Tomozawa contact interaction with a relatively strong coupling to the $K\Lambda$ and $K\Sigma$ channels through the SU(3) structure.

The appearance of the spike-like structure in $\sigma_n/\sigma_p$ is a stable feature resistant to various tests of the model, such as the removal of the genuine resonance states, the inclusion of the higher-order baryon pole term in the loop contribution (cf. Fig.~\ref{fig:diagrams}), or the inclusion of the $\pi\pi N$ channel. 
Also, this spike-like structure is already present at NLO in the chiral expansion, where the triangle graph with the Weinberg-Tomozawa term contributes.

In summary, while the present study does not rule out other explanations of the peak in $\sigma_n/\sigma_p$, such as narrow resonances, it shows the stability of the result of our model, the consistency with other reactions, and delivers a simple and quantitative explanation.

\vspace*{0.3cm}

\noindent {\bf Acknowledgements:} 
This work is supported by DFG\\ (Deutsche Forschungsgemeinschaft, Gz: DO 1302/1-1) and the COSY FFE grant No. 41445282  (COSY-58). We thank I. Jaegle and B. Krusche for discussions on the implementation of the Fermi motion and J.~Haidenbauer for a critical reading of the manuscript.


\end{document}